\documentstyle[11pt,aaspp4,psfig]{article}   

\slugcomment{Accepted for publication on A\&A, 2000 Jan 6th}

\lefthead{Belloni T. et al.}

\begin{document}

\def\spose#1{\hbox to 0pt{#1\hss}}
\def\lta{\mathrel{\spose{\lower 3pt\hbox{$\mathchar"218$}}
     \raise 2.0pt\hbox{$\mathchar"13C$}}}
\def\gta{\mathrel{\spose{\lower 3pt\hbox{$\mathchar"218$}}
     \raise 2.0pt\hbox{$\mathchar"13E$}}}
\def\Msun{{\rm M}_\odot}
\def\msun{{\rm M}_\odot}
\def\Rsun{{\rm R}_\odot}
\def\Lsun{{\rm L}_\odot}
\def\half{{1\over2}}
\def\RL{R_{\rm L}}
\def\zs{\zeta_{s}}
\def\zR{\zeta_{\rm R}}
\def\dJJ{{\dot J\over J}}
\def\dMM{{\dot M_2\over M_2}}
\def\tKH{t_{\rm KH}}
\def\eck#1{\left\lbrack #1 \right\rbrack}
\def\rund#1{\left( #1 \right)}
\def\wave#1{\left\lbrace #1 \right\rbrace}
\def\dd{{\rm d}}

\def\stateA{{A\,}}
\def\stateB{{B\,}}
\def\stateC{{C\,}}

\title{A model-independent analysis of the variability of GRS~1915+105}
\author{
	T.~Belloni\altaffilmark{1,2},
	M.~Klein-Wolt\altaffilmark{1},
	M.~M\'endez\altaffilmark{1,3},
	M.~van der Klis\altaffilmark{1},
	J.~van Paradijs\altaffilmark{1,4}
       }

\altaffiltext{1} {Astronomical Institute ``Anton Pannekoek'',
       University of Amsterdam and Center for High-Energy Astrophysics,
       Kruislaan 403, NL-1098 SJ Amsterdam, the Netherlands}

\altaffiltext{2} {Osservatorio Astronomico di Brera,
       Via E. Bianchi 46, I-23807 Merate (LC), Italy}

\altaffiltext{3}{Facultad de Ciencias Astron\'omicas y Geof\'{\i}sicas,
       Universidad Nacional de La Plata, Paseo del Bosque S/N,
       1900 La Plata, Argentina}

\altaffiltext{4}{Physics Department, University of Alabama in Huntsville,
       Huntsville, AL 35899, USA}

\begin{abstract}

We analyzed 163 observations of the microquasar GRS~1915+105 made with the
Rossi X-ray Timing Explorer (RXTE) in the period 1996-1997. For each
observation, we produced light curves and color-color diagrams.
We classified the observations in 12 separate classes, based on their
count rate and color characteristics. From the analysis of these classes,
we reduced the variability of the source to transitions between three
basic states: a hard state corresponding to the non-observability of the
innermost parts of the accretion disk, and two softer states with a
fully observable disk. These two soft states represent different temperatures
of the accretion disk, related to different local values of the accretion
rate. The transitions between these states can be extremely fast.
The source moves between these three states following certain patterns
and avoiding others, giving rise to a relatively large but limited number
of variability classes. These results are the first step towards a linking
of the properties of this exceptional source with standard black-hole
systems and with accretion disk models.

\end{abstract}


\section{INTRODUCTION}

GRS~1915+105 was discovered in 1992 with WATCH (Castro-Tirado, Brandt \&
Lund 1992) as a transient source with considerable variability (Castro-Tirado
et al. 1994). It was the first Galactic object to show superluminal
expansion in radio observations (Mirabel \& Rodr\'\i guez 1994).
The standard interpretation of this phenomenon in terms of relativistic
jets (Rees 1996) placed the source at a distance D\,=\,12.5 kpc, with a
jet axis at an angle $i$=70$^\circ$ to the line of sight (Mirabel \&
Rodr\'\i guez 1994, Rodr\'\i guez \& Mirabel 1999, but see
Fender et al. 1999).
A counterpart has been found in the infrared (Mirabel et al. 1994), but
the high Galactic extinction prevents the detection of an optical
counterpart.
Before the launch of the Rossi X-ray Timing Explorer (RXTE),
not many observations of this source existed in X rays:
the WATCH and SIGMA instruments on board GRANAT and BATSE/GRO showed
once again that it was very variable in the X-ray band (Sazonov
et al. 1994, Paciesas et al. 1996).

Since the launch of RXTE, the source has been monitored continuously with the
All-Sky Monitor (ASM) instrument, which showed that the 2-10 keV X-ray flux
of the source is extremely variable, unlike any other known X-ray source
(see Remillard \& Morgan 1998). GRS~1915+105 was observed extensively with the
Proportional Counter Array (PCA) on board RXTE: from these data,
the variability of the source appears even more extreme (Greiner, Morgan \&
Remillard 1996). Quasi-periodic oscillations with frequencies ranging from
0.001 to 67 Hz have been observed (Morgan, Remillard \& Greiner 1997, Chen,
Swank \& Taam 1997). A subset of observations, showing a remarkably regular
behavior, has been presented by Taam, Chen \& Swank (1997) and Vilhu \&
Nevalainen (1998), with different interpretations.
Other observations did not show spectacular variability (see Trudolyubov,
Churazov \& Gilfanov 1999).
The source is suspected to host a black hole because of
its similarity with the other Galactic superluminal  source GRO~J1655-40
(Zhang et al. 1994), for which a dynamical estimate of the mass is available
(Bailyn et al. 1995) and because of its very high X-ray luminosity.

Belloni et al. (1997a,b) showed, from the analysis of the energy spectra
of selected observations, that the large variability of the X-ray flux
of GRS~1915+105 can be interpreted as the appearing/disappearing  of
a detectable
inner region of the accretion disk, caused by the onset of thermal-viscous
instabilities. From the comparison of the X-ray color-color diagrams of all
observations available until then, they showed that all
observations were consistent with this interpretation,
with the exception of a few, whose character could not be accounted for.
Additional spectral analysis has been presented by Markwardt, Swank \&
Taam (1999) and Muno, Morgan \& Remillard (1999),
who analyzed in detail the connection between QPOs and
X-ray spectral distribution in GRS~1915+105.

Quasi-periodic variability in the radio and infrared bands has been
discovered (Pooley 1995, Pooley \& Fender 1997, Fender et al. 1997).
Fender et al. (1997) suggested that these oscillations correspond to small
ejections of material from the system. Indeed, these oscillations have
been found to correlate with the disk-instability as observed
in the X-ray band (Pooley \& Fender 1997, Eikenberry et al. 1998,
Mirabel et al. 1998, Fender \& Pooley 1998).
This strongly suggests that (some of)
the matter from the inner disk is ejected in form of a jet throughout the
instability phase.

GRS~1915+105 displays a bewildering variety of variability modes (see
Greiner, Morgan \& Remillard 1996; Belloni et al. 1997a,b;
Chen, Swank \& Taam 1997; Nayakshin, Rappaport \& Melia 1999,
Muno et al. 1999). 
In this paper, we present a classification of the different types of
variability observed in the X-ray emission of GRS~1915+105 and we demonstrate
that all the variations can be accounted for in terms of
switching between just three main spectral states.
All the observations are found to follow this characterization, which
is model-independent. The structure of the paper is the following.
In Section 2, we present the extraction techniques and the tools used
for the analysis. In Section 3, we proceed to the classification of the
observations and present the evidence for the three basic states. In
Section 4, we discuss these results in terms of physical models, in
particular in connection with the disk-instability model presented in
Belloni et al. (1997a,b). A forthcoming paper will include detailed
spectral modeling for a precise (and model dependent)
characterization of these states.

\section{DATA EXTRACTION AND COLOR-COLOR DIAGRAMS}

We analyzed all observations of GRS~1915+105 made between
January 1996 and December 1997 marked {\tt PUBLIC} in the RXTE TOO archive
(at heasarc.gsfc.nasa.gov), plus observation sequences
20187-02-01-00, 20187-02-01-01 and 20187-02-02-00, for a total
of 163 observations. A list of the observations can be found, subdivided in
classes (see below), in Table 1. Each observation consists of one or more
continuous intervals of data (called observation intervals) separated
by earth occultations. Each observation interval has a maximum duration
of $\sim$1 hour
per interval. Our database contains a total of 349 observation intervals.
For each observation interval we produced light curves with 1~sec time
resolution in three PHA channel intervals: A:0-13 (2--5 keV), B:14-35 (5--13
keV), C:36-255 (13--60 keV) (energy conversion corresponding to PCA Gain
epoch 3). We subtract a constant background level of 10, 20 and
100 cts/s in the A, B and C bands respectively, as determined from the
analysis of typical background spectra. Although this is only an approximation,
the high source count rate minimizes the effect of background variations.
From these light curves we produced a total light curve (R=A+B+C) and two
X-ray colors: HR$_1$=B/A and HR$_2$=C/A.

The analysis described below is based on different combinations of these
three parameters, R, HR$_1$ and HR$_2$. The choice of the definition for
the two X-ray colors was made to ensure the linearity of the resulting
color-color diagram (CD): any linear combination of two spectral models
lies on the straight-line segment connecting their locations in the CD.

The energy spectra of GRS~1915+105 are usually fitted with a model
consisting of the sum of a power-law and a multicolor disk-blackbody,
modified by interstellar absorption (see e.g. Belloni et al. 1997b).
This model has become standard for the fitting of spectra of black-hole
candidates.
Figure 1 shows a CD with the location of the two models for different
values of their parameters (photon index $\Gamma$ and temperature
at the inner radius kT$_{in}$).
The interstellar absorption has been fixed to $6\times 10^{22}$cm$^{-2}$,
as determined from spectral fitting (\cite{bel97b,mar99}).
With our choice of colors, HR$_2$ is mostly sensitive to
changes in the slope of the power law component and HR$_1$ to changes in the
temperature of the disk component for the ranges shown in Fig. 1.
The additional presence of an iron line or edge, although significant
for the $\chi^2$ fitting of energy spectra, does not modify the colors
much compared to the overall range covered by the source.
It is important to note that the color analysis is by itself independent
of a specific model. Therefore, although we interpret the results in the
framework of a power law plus disk model, all the observed color
variations are model independent.

In addition to CDs, we also produced Hardness-Intensity Diagrams (HIDs),
where HR$_2$ is plotted versus the total count rate. Although these diagrams
contain absolute flux information and depend strongly on the long term
variability of the source, in some case they are useful for characterizing the
behavior during a single observation.

\section{DATA ANALYSIS}

\subsection{Selection of observations (classes)}

Despite the extraordinarily complex variability displayed by
GRS~1915+105, the
examination of light curves and CDs shows that many features repeat
in different observations. There are cases of observation intervals
more than one year apart which are virtually indistinguishable
from each other. We found that it was possible to classify our 349 observation
intervals into only 12 classes,
based on the appearance of light curves and CDs
(see Table 1). This empirical classification is intended as a first step to
reduce the complexity of the amount of available data,
as well as a way to present the overall picture of timing/spectral
variability of GRS~1915+105 in a model-independent fashion.
We outline the 12 classes  below. One example for each class
(light curve and CD) is shown in Figure 2.

\begin{description}
%
%
\item[$\bullet$ class $\phi$]
Two types of observations are relatively straightforward to separate from the
others: those displaying no large amplitude variations
(less than a factor of two)
nor obvious structured variability (only random noise is seen in light curves
binned at 1 second).  However,
examination of their light curves and CDs shows that these can be further
subdivided into two separate classes:
The first, class $\phi$, is characterized by
a count rate of $\sim$10 kcts/s or less and by a particular position
in the CD, namely HR$_1\sim$1, HR$_2\sim$0.05 (see Fig. 2a,b).
%
%
\item[$\bullet$ class $\chi$]
The second, class $\chi$, has a more varied count rate, ranging between
3400 and $\sim$30000 cts/s (for 5 PCU units) and it
occupies a position in the CD which can vary but is always at HR$_2>$0.1
(see Fig. 2c,d). The cloud of points in the CD is usually diagonally elongated.
%
%
\item[$\bullet$ class $\gamma$]
There are observations that are relatively quiet (class $\gamma$), apart from
quasi-periodic oscillations with a typical time scale of $\sim$10 s and/or
the presence of sharp 'dips' with a typical duration of a couple of seconds
(see Fig. 2f).
In the CD, the quasi-periodicity results in a diagonally elongated distribution
of points. The points corresponding to the dips are separated
from the main distribution, and are located on its left along the
direction of the elongation (See Fig. 2e).
%
%
\item[$\bullet$ class $\mu$]
These are observations that, although showing large amplitude (more than
a factor of two) variability,
show a single `branch' of points in the CD (class $\mu$, Fig. 2g,h).
This branch is elongated diagonally and it is curved to the right in its lower
part. The light curves show rapid quasi-periodic oscillations (typical
time scale between 10 seconds and 100 seconds), sometimes alternated with
stabler periods of $\sim$100 seconds duration.
%
%
\item[$\bullet$ class $\delta$]
A number of observations show red-noise-like variability, which can exceed
a factor of two in range.
These observations form class $\delta$. Notice that in a number of
observations in this class, there are `dips' in the light curve which,
unlike the rest of the variability, are characterized by a significant
softening in both colors (see Fig. 2i,j). These dips last typically 10-20
seconds and, while the out-of-dip count rate is always above 10000 cts/s,
during the dip it decreases to 10000 cts/s or less.
%
%
\item[$\bullet$ class $\theta$]
A group of observations clearly distinguishable from the others is
class $\theta$ (Fig. 2k,l): the light curve has a
characteristic shape, with `M'
shaped intervals with a typical duration of a few hundred seconds,
alternating with shorter (100-200 seconds) low count rate ($\le$10000 cts/s)
periods.
In the CD, two separate distributions are seen: during the `M' events HR$_2$
is above 0.1, while in the low-rate intervals it is considerably softer.
%
%
\item[$\bullet$ class $\lambda$]
The observation presented in Belloni et al. (1997a) and a few similar ones
constitute class $\lambda$ (Fig. 2m,n): the light curve consists of the
quasi-periodic alternation of low-quiet, high-variable and oscillating parts
described in Belloni et al. (1997a).
In the CD, a C-shaped distribution is evident, with the lower-right
branch slightly detached from the rest, and corresponding to the low
count rate intervals (typically a few hundred seconds long).
%
%
\item[$\bullet$ class $\kappa$]
Very similar to the previous class are observations in class $\lambda$. The
timing structure, as shown by Belloni et al. (1997b), is the same, only
with shorter typical time scales (Fig. 2o,p). In the CD, an additional
cloud between the two branches is visible (see \cite{bel97b}).
%
%
\item[$\bullet$ class $\rho$]
Taam, Chen \& Swank (1997) and Vilhu \& Nevalainen (1998) presented
extremely regular RXTE light curves of GRS~1915+105, consisting of
quasi-periodic `flares' recurring on a time scale of 1 to 2 minutes.
There are differences in the observations presented by these authors,
and for this reason we separate them in two classes.
The first, class $\rho$ (Fig. 2 q,r),
is extremely regular in the light curve, and in
the CD it presents a loop-like behavior (described as ring-like in \cite{vil98},
where data with lower time resolution were considered).
%
%
\item[$\bullet$ class $\nu$]
There are two main differences between observations in this class and those
of class $\rho$. The first is that they are considerably more irregular
in the light curve, and at times they show a long quiet interval, where the
source moves to the right part of the CD (see Fig. 2s,t). The second is that,
at 1s time resolution, they show more structure in the profile of the
`flares', notably a secondary peak after the main one (see Fig. 17b).
%
%
\item[$\bullet$ class $\alpha$]
Light curves of observation intervals of class $\alpha$ show long ($\sim$1000
s) quiet periods, where the count rate is below 10000 cts/s,
followed by a strong ($>$20000 cts/s) flare and a few 100s of seconds of
oscillations(see Fig. 2u,v). The oscillations start at a time scale of
a few dozen seconds and become progressively longer.
This pattern repeats in a very regular way. In the CD, the
quiet periods result in elongated clouds, the oscillations in small rings
(not clearly visible in Fig. 2v)
like those of classes $\rho$ and $\nu$, and the flare as a curved trail
of soft (low HR$_2$) points (Fig. 2u).
%
%
\item[$\bullet$ class $\beta$]
This class shows complex behavior in the light curves, some of which
can be seen within other classes. What identifies class $\beta$ however,
is the presence in the CD of a characteristic straight elongated branch
stretching diagonally.
\end{description}

The number of the classes presented above could be reduced, given the strong
similarities between them, but our goal is not to have as few classes as
possible, but to give as comprehensive a description of the source
behavior as possible, in
order to look for basic `states' of the source.
For this purpose, defining a relatively large number of classes means
we are being conservative in order not to overlook important details in
source behavior.
All observation intervals in the sample considered
for this work are covered by this classification,
but it is quite possible that future observations would
require yet other classes.
Some of our observation
intervals can be seen as boundary cases between two classes.
Therefore, our classification is not intended to exhaustively list
mutually exclusive modes of behavior for GRS~1915+105
as: (i) transitions between some classes exist,
(ii) a smaller number of classes would probably be sufficient
to describe our observations, and (iii) more classes probably exist.
The point of our work will instead be to demonstrate that this very
complex behavior in fact follows a few very simple ``universal laws''.
Summarizing, in Fig. 3 we show a histogram with the ``occupation times''
of the different classes in our sample. Noice that class $\chi$ is
by far the most common.

\subsection{Classes $\lambda$, $\kappa$ and $\theta$: the basic states}

Two observations representing classes $\lambda$ and $\kappa$,
I-38-00 (Interval \#3) and K-33-00 (Interval \#2) respectively (notice the
shortened naming convention, explained in the caption to Table 1),
have already been presented by Belloni et al. (1997a,b).
For a better understanding of what follows, we will briefly summarize their
main result, restated using the terminology that we will use throughout
the rest of this work.
Let us start with examining class $\lambda$.
The total light curve, the CD and the HID are shown in Fig. 4.
Panels (a), (b) and (c) in Fig. 4 are the HID, the light curve
(restricted to the points used for CD/HID) and the CD respectively.
The complete light curve of observation interval \#3 is shown in panel (d).
Belloni et al. (1997a) identified  in this observation (and then extended
to all observations)
three types of variability (``states"): a quiescent state at a relatively
low count rate, an outburst state at high count rates, and a flare state in
which the flux shows rapid alternations between the two.
Hereafter, we will call the quiescent and outburst
state \stateC and state \stateB respectively, and we will represent the
corresponding points in all CDs, HIDs and light curves with
squares and circles respectively.
An example of these two states is clearly visible in Fig 4. These are
the two main states corresponding to the absence and presence of a detectable
inner
part of the accretion disk (\cite{bel97a}). The points corresponding to the
two states cluster in two well separated regions of the CD and HID. In the
CD, state \stateB lies in the upper left part of the diagram, while
state \stateC is located more towards the lower right,
at a higher HR$_2$ and lower HR$_1$.
The flare state was interpreted by Belloni et al. (1997a) as oscillations
between the other two, not as a separate state. An
example of this is shown in Fig. 5, where the CD and HID for a ``flaring''
interval of the same observation are shown. It is evident that, although the
high-flux points (state B, circles)
are indeed located in roughly the same region of the CD as before,
the low-flux points (state C, squares)
are considerably softer than those from the long state \stateC
intervals in Fig. 4). From their
analysis of observation K-33-00, Belloni et al. (1997b) showed that
the position of the state \stateC points in the CD depends on the length of
the low-flux interval: the shorter the state \stateC event is, the softer it is and
the more it will be shifted to the left in the CD. Since the
oscillations shown in Fig. 4 have a time scale considerably shorter
than those in Fig. 2, a softening of the state C spectrum occurs.
The identification of these parts of the oscillations with state C is
strengthened by the fact that during them 1-10 Hz QPOs are seen (see
Markwardt et al. 1999; Muno et al. 1999).

On the basis of
this analysis, Belloni et al. (1997a,b) tentatively concluded
that the variability of GRS~1915+105 consists of oscillations between two
separate states: state \stateB corresponds to an accretion disk extending all
the way down to the innermost stable orbit and therefore appears always with the
same spectral colors, while state \stateC corresponds to different portions of
the inner disk being unobservable, and therefore can be found in different
regions of the CD, but always at lower count rate and always
separated from state \stateB.
However, Belloni et al. (1997b) also noticed that,
although this characterization
could be applied to most observations available at the time, one type
of observation could not be reduced to these two states. This type
observation is class $\theta$ in our classification of light curves
(see Section 2). A typical observation of this class is shown in Fig. 6.
The light curve consists of M-shaped intervals reminiscent of the long
state \stateC segments in Fig. 4. (although at a much higher flux level),
alternated with low-rate stretches. In the middle of these low-flux stretches
there are some `flares'. Comparing the CD in Fig. 6 with that in Fig. 4,
we can see that it is indeed plausible
that the M-shaped intervals are consistent with being state \stateC
(at a slightly higher HR$_1$ and much higher count rate), and
that the low-flux points cannot be identified with
state \stateB, being much softer.
Since these soft points cannot be assigned to either of the two
states, we define a new state (state \stateA). In all the CDs,
the points corresponding to this state will be marked with diamonds.
During the `flares' in the low-flux intervals, in the
CD the source moves back almost to the state \stateC points.

State \stateA is characterized by consisting of low-flux, soft intervals, as
opposed to state \stateC, which is low-flux and hard. The presence of this
additional state is very evident in an
observation
like that shown in Fig. 6, but in retrospect it could of course
be present also in other observations.
Going back to Fig. 4, we notice that at the end of a long state \stateC period
(see Fig. 4d) there is a short ($<$10s) dip. This dip corresponds to soft
points in the CD (diamonds in Fig. 4c) which at the bottom of the dip reach the
same position as those in Fig. 6. We find that in observations
of class $\lambda$,
this dip is {\it always} present as a transition between state \stateC
and state \stateB, not only at the end of the long state C intervals,
but also during the much more rapid flaring.
By examining in detail the oscillations shown in Fig. 5, one can see that
indeed a soft dip lies in between all \stateC-\stateB transitions, although
the position in the CD in this case is much higher and in some cases it
overlaps with that of state \stateB (this overlap will be discussed
extensively below). Also, the dip in the light curve corresponding to state
A is not as deep as in the previous case.

Moving to observation K-33-00 (class $\kappa$), an examination of a
short sample of the light curve shows that, although the time scales
are shorter, the source behaves like in the $\lambda$ case
(see Fig. 7). This was already recognized by Belloni et al. (1997b), who
based on these data their analysis of the disk instabilities for states
B and C, but it also applies when state A is considered.

\subsection{The main color loop}

It is useful to summarize the results from the first three classes
before presenting more data.
In the three observations we have discussed so far, we could identify
three basic states:

\begin{itemize}

\item {\it \stateB}: high rate, high HR$_1$;
\item {\it \stateC}: low rate, low HR$_1$, variable HR$_2$ depending on the
	length of the event;
\item {\it \stateA}: low rate, low HR$_1$ and HR$_2$

\end{itemize}

As far as the transitions between the three states go, in two cases
(classes $\lambda$ and $\kappa$), the source invariably follows the
sequence: \stateB-\stateC-\stateA-\stateB and so on. In the third case
(class $\theta$), there are only oscillations between two states
(\stateC and \stateA). No direct transition from state \stateC
to state \stateB is observed. A schematical representation of this can be
seen in Fig. 8.
Notice that in some observations, state \stateA is not completely
separated from state \stateB in the CD: this blending between the
two classes is discussed extensively in the following sections.
Before discussing this diagram however, we must
go through the other classes.

\subsection{Class $\chi$: state \stateC only}

As described before, class $\chi$ is characterized by the absence of
strong variability and HR$_2>$0.1. A large number of observation intervals
belonging to this class correspond to observations taking place during
three periods when no strong variability was observed on long (hours to months)
time scales with the RXTE/ASM.
The three periods are identified in Fig. 9 by arrows.
Although the two short such intervals are well defined in the ASM light curve
(MJD 50280-50311 and MJD 50730-50750 respectively),
the boundaries of the long one are not so clear. We define its start as the
time after which {\it all} PCA observation intervals show no structure in the
1s light curves, and its end as the time after which
structure in the time variability is observed again in the PCA. This leads
to the following start-end times: MJD 50410-50550, corresponding to
observations K-04-00 to K-21-01. Observation K-03-00, the last before this
period, shows a clear quasi-periodic structure corresponding to class
$\rho$, and observation K-22-00, the first after this interval, shows
again a quasi-periodic structure, but classified as class $\alpha$.
Consistently, we applied the same method to re-define the time boundaries
of the two shorter intervals, in order not to limit ourselves to the flat-bottom
part of the light curves. We obtained the following time intervals:
MJD 50275-50313 and MJD 50729-50750 respectively.
We subdivided the observations of class $\chi$ into four
sub-classes: $\chi_1$, $\chi_2$, $\chi_3$ correspond to the three quiet
intervals (in time sequence), while $\chi_4$ comprises the remaining
observation intervals, a number of which are located close to the long
quiet interval. The assumption underlying this division is that the three
long quiet intervals are indeed single intervals of no variability.
Notice that our $\chi_1$ and $\chi_3$ intervals correspond to the
``plateau'' states in Fender et al. (1999).
Figure 10 illustrates a selection of four observations in this class.
It can be seen that the position in the CD is similar to that associated
to state \stateC (see the previous section), and we therefore conclude
that class $\chi$ is an example of state \stateC only. The absence of
flares and visible quasi-periodicities, as well as the presence of
variability in the form of white noise at time scales of 1 second
(see Markwardt, Swank \& Taam 1999
and Trudolyubov, Churazov \& Gilfanov 1999),
further supports this identification as the same state as the low
flux periods described in Section 3.2. However, it is clear that there is
substantial spectral variability between different observations, as the clouds
of points in the CD are found in quite different positions (Fig. 10).
As mentioned in Section 3.2, for shorter events, the position of state \stateC
is also variable.
In order to examine these variations, we computed the average values of the
X-ray colors for each observation and plotted them into the same CD (Fig. 11).
The lines corresponding to some spectral models are included in the plot.
The points (filled symbols in Fig. 11) clearly distribute along three
separate branches. Classes
$\chi_2$ and $\chi_4$ follow a flatter trend than classes $\chi_1$ and
$\chi_3$. From Fig. 11, one can see that large movements to the right in the
CD are likely caused by hardening of the power law component. However, a
simple power law model with the ``standard'' value of
6$\times 10^{22}$cm$^{-2}$ for the interstellar absorption fits only
the $\chi_1$ and $\chi_3$ points.
On the basis of the CD, the other points (classes $\chi_2$ and
$\chi_4$) can be fitted by a power law
model with a lower absorption (2$\times 10^{22}$cm$^{-2}$).
An absorption of 2$\times 10^{22}$cm$^{-2}$ is compatible also
with the $\chi_1$ and $\chi_3$ points only if an additional
high-energy cutoff at $\sim$5 keV is included (see lines in
Fig. 11).
Detailed spectral fits, which will be presented in a
forthcoming paper, indeed indicate that both such a reduced
absoption and a high-energy cutoff are present.
Notice that 6$\times 10^{22}$cm$^{-2}$
is the standard value adopted for GRS~1915+105, while the lower value
2$\times 10^{22}$cm$^{-2}$ is still consistent with the galactic N$_H$
as determined from radio measurements (Dickey \& Lockman 1990).
In this scenario, a large intrinsic absorption is present
in the system and would be variable on a time scale longer than that of
the long quiet intervals examined here.

\subsection{Class $\phi$: state \stateA only}

Class $\phi$ is the second of the two classes where no variability is observed.
An example of light curve and CD is shown in Fig. 12 (upper panels). Compared to
class $\chi$, there is less white noise in the light curve (although in some
cases red-noise variability is observed), and the
points in the CD are much softer and more concentrated. The net count
rate is $<$10 kcts/s. To determine whether this class corresponds to
state \stateA or state \stateB, we need to compare the CD with that of
an observation which shows both states. In Fig. 11 we show the average
colors of all observations in class $\phi$ (see Table 1) together with the
state \stateB points from Fig. 4. By comparing Fig. 4c and Fig. 11,
it is evident that observations in class $\phi$ have to be identified
with state \stateA. Note that the quietness of the light curve is also similar
to the low-rate state \stateA points from class $\theta$ (see Fig. 6).
From Fig. 11, one can notice a continuity between state B and state A
points, suggesting that the two states might not be completely different
but rather part of the same state. This will be discussed extensively below.
\subsection{Class $\delta$: \stateB and \stateA}

An example of ligh curve, CD and HID for an observation in class $\delta$
is shown in Fig. 12 (lower panels). The light curve shows considerable
`flaring' variability.
These variations are not accompanied by changes in X-ray colors, with
the exception of a number of easily identifiable 'dips', which correspond
to a spectral softening. From the CD, we can identify the dips as
occurrences of state \stateA, while the rest of the time the source is
in state \stateB. It is noticeable that these state \stateA events are
the only ones when the source approaches or becomes weaker than 10 kct/sec.

\subsection{Class $\mu$}

Observations from class $\mu$ show, like some parts of observations in class
$\lambda$, large flaring
variability and a characteristic shape in the CD: it is elongated diagonally
and bends towards higher HR$_2$ values
in its lower part (see Fig. 13).  In panels b, c and d from this figure the part
of the light curve
used to produce the CD, the CD and the total light curve are shown
respectively. When we compare
this figure to Fig. 5, where the flaring state of observation I-38-00 (Orbit 3,
class $\lambda$) is
shown, the striking resemblance between the two is evident. We can therefore
identify state C in
both the light curve and the CD (Fig. 13). Its relatively soft position in the
CD compared to the
position of state C in observation I-38-00 (Orbit 3) as shown in  Fig. 4c, can
again be explained by
the short time scales observed (see section 3.2). Although, from the light curve
in Fig. 13b (and
comparing this with the one shown in Fig. 5b) state A and B are evident, they
are not easily
identified in the CD (Fig. 13c). The two states seem to be spectrally
identical, as they take the
same position in the CD. In order to check whether there is indeed a difference
between state A and
B, we produced a HID from the count rate and the HR1
values, again using only the part of the light
curve shown in Fig. 13b. From this HID, shown in Fig. 13a, and the light curve
in Fig. 13b, we
see that in this class state A is also present, although its major difference
with state B is count rate. State A and C
both have count rates between 10 and 25 kcts/s, and only in state B the source
becomes higher than
30 kcts/s.

Besides the rapid state alternations, stabler periods where the count rate
remains at a high level
around 40 ckts/s are observed in some light curves in class $\mu$ (see Fig. 2h).
During these intervals hardly any spectral variation is seen, and the source
takes a position in the CD with relatively soft HR2
(below $\sim$0.12) and hard HR1 (above $\sim$1.0) values. We therefore identify these
intervals as periods
were the source remains in state B for a extended period of time. They can be
compared with the
state B intervals in Fig. 4. Although at a higher count rate, the position in the
CD is the same.

\subsection{Variations within state A}

As we have seen, in some cases examined so far the distinction between classes A
and B in the CD becomes uncertain, complicating the simple picture sketched in
Fig. 8. This is summarized in Fig. 14a, where we plot
the CD from state B in observation I-38-00 (Orbit 3) and the average CD from
class $\phi$, as already shown in Fig. 11, together with the dips of all state A
points shown in the previous sections. Indeed, most of state A points
are clearly separated from those of state B, but some are not and a couple
of them are even harder than all state B points shown.
However, if the same points are plotted in a HID (see Fig. 14b), those
`peculiar' state A points are immediately distinguishable from the state~B ones.
What can be seen in Fig. 14b is that the colors of both state A and state B
points are correlated with the count rate, but follow different branches.

From Fig. 14, we can re-define state A in the following terms:
\begin{enumerate}
\item in the light curve, a sharp dip, with a typical transition time scale
	between out-of-dip and dip of a few seconds;
\item at low count rates (below ~15 kcts/s),
   the position in the CD is relatively soft (HR$_1<$1.1) and separated
   from state B;
\item at higher count rates, the position in the CD overlaps with that of
   state B, but follows a different branch in the HID, located at lower rates.
\end{enumerate}

This branch will be examined in more detail below.

\subsection{Class $\gamma$}

In Fig. 15 we show observation K-39-00 (Orbit 1), representing class $\gamma$.
All observations from this class have a count rate centered around
$\sim$20kcts/s, and show evidence of quasi-periodic oscillations with
period $\sim$60-100 s (see Fig. 15d), in some cases with an amplitude
as high as a factor of two (see
Greiner, Morgan \& Remillard 1996 and Morgan, Remillard \& Greiner 1997).
Simultaneously, faster $\sim$10 s QPOs are
observed (see Fig. 15b). A number of observations show also sharp dips
as the ones observed in class $\delta$. These dips always occur in
correspondence to a minimum in the low-frequency oscillations.

In the CD (Fig. 15c), by comparison with previous cases, it appears that the
oscillations belong to state B and the dips to state A (compare Fig. 15c
with Fig. 14a).

\subsection{Class $\rho$ and $\nu$}

Observations in class $\rho$ have been presented by
Vilhu \& Nevalainen (1998), although limited to low time resolutions
(16 seconds). The light curve consists of a very regular and characteristic
pattern, which repeats on a time scale between one and two minutes (see Fig.
16). As noticed by Vilhu \& Nevalainen (1998), these regular variations
are reflected in a clockwise loop in the CD (Fig. 16). Notice that at a
resolution of 1 second, the CD looks much less circular than in Vilhu \&
Nevalainen (1998), who used a resolution of 16 seconds, clearly not
sufficient for the analysis of these observations. A comparison with
the classes examined in the previous sections clearly indicates that
the loop is constituted by oscillations between state B (high count rate)
and state C (low count rate), without a clear evidence of a state A.
However, when examining the light curves at a higher time resolution,
a complex type of variability is resolved (Fig. 17).
The high count rate part of the cycle shows sharp dips, with count rate
changes of a factor of 3 in less than one second. The overall structure
of this variability does not repeat between different events,
with the only exception of the small peak after the last dip. The two
examples in Fig. 17 are shifted so that this peak corresponds to time 0.
At such high time resolution
it is difficult to follow a few points in the CD, since the count rate
in the high-energy band becomes uncertain. From the observation in Fig. 16
and 16, binned at 1/8 s, we considered only the points in the 10 seconds
preceding the last peak (in order to isolate the high count rate sections),
and averaged the colors of the points with a total rate $<$12 kcts/s.
The resulting average colors are indicated by the diamond in Fig. 16c:
the point falls in the soft region of the CD plane, corresponding to the
position of state A. All the states are therefore represented in the light
curves of this class. Notice that for a few observations in this class,
the high time resolution light curve is much smoother, but still the dip
preceding the last peak is present.

An observation from class $\nu$ has been presented by Taam, Chen \& Swank
(1997).
In the light curve, class $\nu$ looks rather similar to the previous one
(see Fig. 18). The main differences with class $\rho$
are the possible presence of long quiet intervals between
trains of oscillations (not shown in the example considered here, but see
Fig. 2),
the different characteristic shape in the 1s light curve, and the
lower regularity of the oscillations. The very fast dips observed in class
$\rho$ can also be seen here, but are less deep. The long quiet interval,
followed by a short high flux flare sometimes observed in observations
included in this class (see Section. 3.1) is very similar
to that in class $\alpha$ and $\lambda$ so it will not be discussed here.
The difference in the regular light curve of class $\nu$ with respect
to class $\rho$ consists in the presence of a dip at the end of
the high-rate part of the cycle, which creates a secondary peak.
This is reminiscent of the dip which gives origin to the small peak
in class $\rho$, although in this class the dip and the peak are much longer,
being both a few seconds long.
By looking at the CD, one can see that the smooth clockwise movement
from class $\nu$ is here interrupted during its ascending part by the dip,
which is very soft. After the dip, the source moves to the hard part of
the diagram. Following the structure of the previous sections, we can
identify the long slow rise in the light curve with state C, followed by
state A (the dip) and state B (the recovery from the dip, showing up as a
secondary bump). Notice that in this class, the typical colors of both
state B and state A are considerably harder than in most of the cases
described in the previous sections.

\subsection{Class $\alpha$}

The light curves from observations belonging to class $\alpha$ show a
repetitive pattern of long ($\sim$ 1000 s) quiet intervals and periods were a
strong (up to $\sim$ 30 ktcs/s) flare is followed by a sequence of smaller
oscillations (see Fig. 19d). The length of these oscillations increases with
time and often their peak intensity decreases.
The quiet intervals mentioned above can easily be identified with
state C. They show the characteristic shape in the CD (Fig. 19c) and
in the light
curve(Fig. 19b), also seen in previous cases (see Fig. 4 and 6).
The strong peak following each quiet interval looks practically identical to
the quasi-periodic ones from class $\rho$ (Fig. 16).
Not only do they have the same shape in the light curve
(compare Fig. 19b and 19e with Fig. 16b), they also show the same
loop in the CD. This can be seen in Fig. 19f, were the CD  belonging to the
peak from Fig. 19e
is shown. Therefore, during this peak the source moves through all
the three states as explained in Section 3.10. In some observations, the
initial strong peak shows a marked HR2-soft `dip' similar to that observed in
class $\nu$ (Fig. 18).
All the smaller oscillations resemble the ones shown
in Fig. 18 for class $\nu$.

\subsection{Class $\beta$}

Observations grouped in class $\beta$ show a large variety of complex
behavior and for this reason are described as last, in order to highlight
the similarities with other classes.  An
example of this can be seen in Fig. 20d and Fig. 20b, were we show observation
K-45-03 (Observation interval \#1).
Periods of large variability (between $\sim$ 8 and $\sim$ 40 kcts/s), similar
to those seen in class $\lambda$
(see Section 3.2), are alternated by quiet intervals. These quiet intervals
are similar to those from
observations in class $\theta$, the main difference being the fact that
here the dip phase after the
strong peak is much less deep. The resulting CD (Fig. 20c) is characterized
by a long elongated cloud
stretching diagonally, and it shows a great resemblance with the CD's seen in
class $\lambda$ (Fig. 4).
In order to identify the different states in the observation shown in Fig. 20,
we make use of the similarities
with the other classes, as mentioned above. From comparison with class $\theta$
(Fig. 6), we can
therefore identify state C and A in the quiet interval, as indicated
in Fig. 20b. The
positions in the CD for these states are the same as found in class $\theta$
(see Fig. 6c): a
relatively soft state A (below HR1=1.0 and HR2=0.1) and a state C which is much
harder in HR2 ($\sim$0.12).
As we have seen in the previous sections, after state A the source
can make a fast transition into state B, as seen for instance in class
$\lambda$ (Fig. 4b). However, in the case presented here the transition to
state B is much more
gradual. The spectrum becomes progressively harder, resulting in a connection
between state A and B in the CD (the diagonal ``finger-like'' structure).
The quiet interval therefore contains all the three states A,B and C. As
mentioned before, the periods
showing the large variability are rather similar to the ones shown in class
$\lambda$ (Fig. 5). Only
here the large dips all have the same count rate ($\sim$ 8 kcts/s, see Fig.
20b,d), whereas in class
$\lambda$ not only the average count rate in the dips is higher ($\sim$ 12
kcts/s), but also
different count rates are observed there (Fig. 5b). Nevertheless we
can conclude that
during these large variations, as for the ones seen in class $\lambda$, the
source is switching between
al the three states A,B and C. This can be seen by comparing Fig. 5b with Fig.
20e, were a part of the
variation interval is shown. The only difference is that in Fig. 20e the
state A intervals have a lower count rate than the state C ones.
As a result of this oscillating between states during these
variations, two branches for state B appear in the HID (Fig. 20a). When the
source moves from state A
to state B it follows the branch on the right in the HID, but the
transition from state B back
to A is characterized by a softer HR1 ($\sim$1), thus following the branch on
the left. It is also
possible, as explained before, to make the transition from state B to state C.
The source then again
follows the left branch, but makes a turn to higher HR2 values as it moves to
state C. The presence of two branches corresponding to state B will be
discussed in the next Section.
Notice that class $\beta$ is the one that has most extensively been studied,
especially in multi-wavelength campaigns (see Markwardt, Swank \& Taam 1999;
Eikenberry et al. 1998). It is clearly the most complex and possibly the
most important of all our 12 classes, and it is presented last here only
because in order to understand its structure one needs to examine simple
classes first. Its analogies and differences with class $\lambda$ are very
important, but a complete analysis of them is beyond the scope of this
paper.

\section{DISCUSSION}

\subsection{The disk-instability model: states B and C}

In the previous sections, we classified and described in detail the whole
complex phenomenology displayed by GRS~1915+105 in the first two years of
RXTE observations. Despite the extreme variety of behavior,
our analysis allowed us to identify three main states, the alternation of
which are at the base of all the light curves and CDs included in Table 1.
While all that has been said before is completely model-independent,
we will now discuss the results in terms of the
``standard'' spectral model consisting of a
disk-blackbody component and a power law.
The three states can be characterized (and identified) in the following way:

\begin{itemize}
\item State A: CD position well above the power-law line
	(i.e., with a substantial
	contribution to the flux by a disk component with kT$>$1 keV).
      Mostly little
	time variability, sometimes red-noise variability.

\item State B: CD position above the power law line, higher than state A.
	Substantial red-noise variability on time scales $>$ 1s.

\item State C: CD position upon or below power law line (either no disk
	contribution or very soft disk inner temperature ($\sim$0.5 keV).
	White noise variability seen on time scales of 1s.
\end{itemize}

If no structured variability is present in an observation (i.e. classes
$\phi$ and $\chi$), state C can be distinguished from the others
by the position in the CD with respect to the power-law line, while
state A and state B differ by their value of HR$_1$. State A has HR$_1<$1.1,
state B has HR$_1>$1.1.
In all the other cases, state A and B can only be separated by comparing the
points in the CD: state A points are either softer or fainter than state B
points.

Notice that a crude subdivision in states, based on a single observation
(K-45-03, class $\beta$) has been presented by
Markwardt, Swank \& Taam (1999). Their
``quiet state'', ``low-frequency noise state'' and ``1-15 Hz state'' can be
identified as specific instances of our A,B,C states.
respectively. Following Belloni et al. (1997a,b), from the position in the CD,
we can identify state B with the typical situation observed in black-hole
candidates in high-flux states: the energy spectrum is the
superposition of a thermal component originating from an accretion disk
which is observable down to the innermost stable orbit, and of a power-law
component with a steep slope (see Belloni et al. 1997a, Markwardt, Swank
\& Taam 1999). With the same spectral model, in state C the power-law
component is much more dominant and the disk component softer.
Let us compare our analysis with proposed models, abandoning for the
moment our model-independency.
In the model by Belloni et al. (1997a,b), this state corresponds to the
unobservability of the inner portion of the accretion disk, due to the
onset of an instability. As the refill time $\tau_{refill}$ is related
to the viscous
time scale at the inner edge R$_{in}$of the observable part of the disk, the
re-fill time for the inner region is
\begin{equation}
\tau_{refill} \propto {\rm R}_{in}^{3.5} \dot {\rm M}^{-2}
\end{equation}
This means that the larger the region affected by the instability, the
longer the state-C event will last. From Belloni et al. (1997b), although
their definition of CD is slightly different from that used in the
present work, we can see that the longer the state-C event, the harder the
X-ray colors. Since in state C, as compared to state B,
the disk component becomes softer, this means that
in state C the power-law component hardens. This effect is observable in our data
too: the longer state-C events are located to the right in the CD (see
Fig. 2). Therefore, our analysis is in agreement with the disk-instability
model, although of course a detailed spectral modeling of all observations
would be needed to say something more firm.

\subsection{Long-lasting effects of the instability}

As we have seen, during the three long quiet intervals included in our
sample, the CD analysis shows that the spectrum of GRS~1915+105 is in
first approximation consistent with a pure power law, possibly with a
high-energy cutoff. All these observations have been identified with
state C, and since they appear in three separate long events, it is
natural to associate each single long quiet period with a single state C
event. In the framework of the disk-instability model applied to GRS~1915+105
it is indeed possible to have a re-filling phase that lasts for a month
or more. This can be achieved with a large radius of the unobservable
inner region of the disk, which as we have seen is associated with a
softening of the disk component and a flattening of the power-law
component. During these periods, the soft component becomes therefore
so soft that it is not detected anymore (due to the high value of
interstellar absorption) and the power-law component becomes progressively
harder (see Fig. 11). However, the instability model applies only to the
inner radiation-pressure-dominated region of a Shakura \& Sunyaev (1973)
accretion disk, which for reasonable parameters cannot extend beyond a few
hundred kilometers from the black hole. This means that in order to
reach the relevant time scales, the accretion rate $\dot {\rm M}$ must also
be lower (notice the strong dependence of $\tau_{refill}$ on $\dot {\rm M}$).
The association between the long quiet periods and single state-C
events is strengthened by the observed properties of the 1-10 Hz QPOs in
both cases (Markwardt, Swank \& Taam 1999, Muno, Morgan \& Remillard
1999; Trudolyubov, Churazov \& Gilfanov 1999).
Muno et al. (1999), in their analysis of a large number of RXTE observations,
find a correlation between the frequency of the 1-10 Hz QPO  and the
inner radius of the accretion disk (as measured through spectral fits).
If during the quiet intervals the disk has a large inner radius (so large
that the disk component is not seen in the PCA spectra), following this
correlation, the QPO frequency should alwasy be rather small, while it can
be as high as 10 Hz (Trudolyubov, Churazov \& Gilfanov 1999). However,
Belloni et al. (1997b) remarked that what can be measured during an
instability phase is the inner disk radius at the beginning of the
instability. The slow refilling is not observable in the energy spectrum,
while the QPO frequency is indeed seen to increase with time as the disk
is refilled. Therefore, what must be associated to a large radius is the
{\it slowest} QPO observed in such an interval, which is about 0.6 Hz
(Reig et al., in preparation). This would not be inconsistent with the
correlation by Muno et al. (1999).

As we have shown in Section 3.4, the differences between different
long quiet intervals can be interpreted as differences in N$_H$ (between
2 and 6$\times 10^{22}$cm$^{-2}$), but also as differences in high-energy
spectral cutoff. A detailed analysis of the energy spectra from these
observations will be reported in a second paper. 
Notice, however, that a cutoff in the broad-band
X-ray spectrum of GRS~1915+105 has been observed by Trudolyubov, Churazov
\& Gilfanov (1999) for the same set of observations, although their value
for the cutoff energy (70-120 keV) appears too high to explain these
differences. The presence of a high-energy cutoff is also reported by Muno,
Morgan \& Remillard (1999) and Feroci et al. (1999).

\subsection{The soft state (state A)}

We have seen that, besides state B and state C, a third state appears
in most of the light curves we have analyzed.  While in most cases
this state is well defined, in some observations a gradual transition
between state A and state B is observed, although the transition {\it to}
state A is always sharp. We consider it a different state, rather than
the softer end of a variable state B because of the
sharp transitions.
Looking at the CDs, we can see that in the disk+power-law model, the
difference between state A and state B could be due to a difference in
power law slope, but only if combined with very specific simultaneous
changes in the inner temperature of the disk. Also, the changes in
power law slope would be very large. A simpler model is to
assume that the inner temperature of the disk changes, since the
state A to state B line moves parallel to the disk-blackbody line (see
Fig. 14). Preliminary spectral fits indicate that indeed there are
variations in the temperature of the disk-blackbody component.
If the inner radius of the disk
does not change, variations in the temperature of the disk are directly
associated with variations in the local accretion rate through the inner
radius. To test whether such local variations can happen on time scales
below a second, since they are connected to the viscous time scale at
the inner edge of the disk, we can take the relation measured by
Belloni et al. (1997b) between viscous time scale and radius, and extrapolate
it down to their minimum observed inner disk radius of $\sim$20 km.
We obtain a time scale of $\sim$0.5 s, therefore of the right order
of magnitude (differences in average accretion rate will influence this
number).

The interpretation of the difference between state A and state B in terms
of a difference in inner temperature of the accretion disk is also in
agreement with the presence of cases where the transition is gradual. In
this framework, the disk changes temperature in response to changes in the
local accretion rate, changes which can be gradual or sudden. What makes
us consider state A a separate state is the presence of the sudden
transitions. In
particular, the transitions {\it to} state A are always fast.

The scenario that emerges from this work is the following. The accretion
disk can have different temperatures, depending on the accretion rate.
The variations in temperature can sometimes be very fast. This is in
principle completely independent of the onset of the instability described
in Belloni et al. (1997a,b), since A--B transitions are observed even in
absence of instability events. Something similar might be observed in other
sources, like in the case of GX~339-4 (Miyamoto et al. 1991).
However, whenever the disk instability is at work, it triggers such
temperature changes, since as we have shown above at the end of an
instability event (when the inner portion of the disk is not observable),
the disk switches {\it always} to state A and {\it never} directly to
state B. This is the only link between the two processes (temperature
changes and instability) and can provide important clues on both
mechanisms.

\subsection{Time variability and the three states}

The detailed time variability of GRS~1915+105 has been studied by a number
of authors (Morgan, Remillard \& Greiner 1997,
Chen, Swank \& Taam 1997, Trudolyubov, Churazov \& Gilfanov 1999,
Markwardt, Swank \& Taam 1999, Muno, Morgan \& Remillard 1999, Feroci et
al. 1999, Cui 1999).
Following these results, it is interesting to compare
the presence of the different QPOs observed in the light curves with the
three states described here.
The so-called ``intermediate'' QPO, usually between 1 and 15 Hz, is
clearly connected to state C (see Markwardt, Swank \& Taam 1999; Muno,
Morgan \& Remillard 1999; Trudolyubov, Churazov \& Gilfanov 1999). As
one can see from these works, this QPO appears always and only during the
state C intervals. Since the frequency of this QPO is proportional to
the observed rate (see Markwardt, Swank \& Taam 1999), it is possible that
the QPO is a tracer of the keplerian motion at the (large) inner radius
of the observable disk, but no precise measurements support this
conjecture.
If the 67 Hz QPO (Morgan, Remillard \& Greiner 1997)
is indeed associated to either the keplerian motion or relativistic
precession at the innermost stable orbit of the accretion disk (Morgan, Remillard
\& Greiner 1997 Zhang, Cui \& Chen 1997), it should
not be observed during state C, when the observable inner radius of the
disk is much larger than the innermost stable orbit.
Although an exhaustive analysis has not
been made, Trudolyubov, Churazov \& Gilfanov 1999 do not report a detection
of such a QPO in the state-C observations they analyzed.
As far as the low-frequency QPOs (in the frequency range well below 0.1 Hz)
are concerned, it is clear from our analysis that they are simply caused
by the state oscillations presented above.

\subsection{How many variability classes?}

As we said above, it is not within the purposes of this work to provide a
complete classification
of all the light curves in terms of the absolute minimum number of possible
classes. However, it is interesting to note that the three states described
in the previous sections and the state transitions between them could in
principle give
rise to a much larger variety of light curves, while only a limited number
has been observed so far. In particular, some of the classes seem to repeat
in an almost indistinguishable manner separated by months or even years.
It is therefore clear that the structure of the time variability, i.e.
the specific alternation of states, is not controlled by a random parameter,
but is related to physical quantities.

\subsection{Relation to standard black-hole candidates}

It is interesting to note that, when in state B, GRS~1915+105 shows
energy spectra and power spectra not unlike those of the so-called
Very High State of black-hole candidates (see van der Klis 1995).
The energy spectrum is
a typical disk component down to a few dozen kilometers radius, plus
a rather steep power-law component. The power spectrum is also similar
to that of a VHS (see Markwardt, Swank \& Taam 1999; Muno, Morgan
\& Remillard 1999;
Belloni 2000a). State A is softer and usually (but not always) less
variable, and shows sometimes sharp transitions with state B: something
similar, although less spectacular, has been observed in the Very High
State of GX~339-4 (the so-called "flip-flops", Miyamoto et al. 1991).
We can conclude that when it is observed in one of these two states,
GRS~1915+105 does not differ much from a standard black-hole candidate.
During state C, an instability occurs, instability which is not yet
observed in any X-ray source, and this is what makes this source
peculiar. Also, there is evidence that the instability is related
to the ejection of superluminal jets from the source (Pooley \& Fender
1997, Eikenberry et al. 1998, Mirabel et al. 1998). Despite
the fact that the physical conditions are rather different, the power
spectra are very similar to those of ``canonical'' black-hole candidates
in their Intermediate state (Belloni 2000a; Trudolyubov,
Churazov \& Gilfanov 1999). The energy spectra are also similar to the
Intermediate State, with a power-law component with a slope of 2.2-2.5
and a soft component (see Belloni et al. 1997b, Markwardt, Swank \& Chen
1999).
The association to the Intermediate State becomes stronger when we
look at the long quiet periods: here the power law becomes flatter and
reaches 1.8 (see Fig. 11 and Trudolyubov, Churazov \& Gilfanov 1999)
and the characteristic frequencies in the power spectrum become smaller
(Trudolyubov, Churazov \& Gilfanov 1999). Both the timing and the spectral
features approach those of the canonical Low State. Since a similar instability
is not observed in standard sources, we can conclude that most likely the
Intermediate State is {\it not} caused by such an instability, but
that the instability mimics the conditions that give rise to an Intermediate
State. In other words, the source looks like it is in a ``canonical''
Intermediate
State because the instability produces the necessary conditions for such
an energy spectrum and such a power spectrum to being produced, i.e. a missing
or invisible inner disk.

\subsection{The twelve classes as standard modes?}

We have shown that it was possible to classify the remarkable variability
of GRS~1915+105 in twelve classes. Although more ``modes'' of variability
might be present, it is remarkable that only a handful of classes can
be used to represent all observations. Indeed some patterns repeat
almost exactly at a distance of months (see Belloni 2000b). Clearly, the
presence of three basic states, even with well defined characteristics,
is not enough to explain these features. Despite the almost infinite
number of ways to alternate the three states, GRS~1915+105 ``chooses''
only a few. This indicates the presence of a small number of modes of
variability, modes which must be connected to basic properties of the
accretion disk.

\section{CONCLUSIONS}

We analyzed a large set of RXTE/PCA observations of GRS~1915+105 by means
of X-ray color diagrams. We could classify the light curves in 12 separate
classes based on the color variability and the light curve. In each of these
classes, we could reduce all source variations to the alternation of three basic
states. We interpret one of the states as the product of a thermal-viscous
instability as proposed by Belloni et al. (1997a,b), and the other
two as reflecting different inner  temperatures of the accretion disk.
This classification provides the necessary background for a detailed
analysis based on spectral fitting.

\acknowledgements

MM is a fellow of the Consejo Nacional de Investigaciones
Cient\'{\i}ficas y T\'ecnicas de la Rep\'ublica Argentina.
This work was supported in part by the Netherlands Organization for
Scientific Research (NWO) under grant PGS 78-277 and the Netherlands
Foundation for Research in Astronomy (ASTRON) under grant 781-76-017.
JVP acknowledges support from NASA under contract NAG 5-3003.

{}

\clearpage

\figcaption[]{Color-color diagram (CD) with the colors defined in the text.
	The two lines correspond to a power law with different photon indices
	$\Gamma$ and to a disk-blackbody model with different inner
	disk temperatures kT$_{in}$. The adopted N$_H$ value used is
	6$\times 10^{22}$cm$^{-2}$. Temperatures increase by steps of 0.1
	keV as the points move up, with the marked point corresponding to
	kT=2.0 keV. Power law indices increase by steps of 0.1 as the points
	move right, with the marked point corresponding to $\Gamma$=2.5.
	}

\figcaption[]{One example light curve and CD from each of the 12 classes
	described in the text. The light curves have a 1s bin size, and
	the CDs correspond to the same points. The class name and the
	observation number are indicated on each panel.}

\figcaption[]{Histogram of occupation time for each of the twelve classes.}

\figcaption[]{Sample observation from class $\lambda$. Panels a,b,c contain
	only part of the data, while panel d shows the whole dataset.
	(a) Color-intensity diagram; (b) light curve; (c)
	color-color diagram; (d) complete light curve. Diamonds, circles and
	squares correspond to states A, B, and C respectively. The vertical
	lines in panel b separate different states.}

\figcaption[]{Same as Fig. 3, but with a different subset of the observation
	selected for panels a,b,c.}

\figcaption[]{Sample observation from class $\theta$. Panels a,b,c contain
	only part of the data, while panel d shows the whole dataset.
	(a) Color-intensity diagram; (b) light curve; (c)
	color-color diagram; (d) complete light curve. The symbols are the
	same as in Fig. 3}

\figcaption[]{Sample observation from class $\kappa$. Panels a,b,c contain
	only part of the data, while panel d shows the whole dataset.
	(a) Color-intensity diagram; (b) light curve; (c)
	color-color diagram; (d) complete light curve. The symbols are the
	same as in Fig. 3}

\figcaption[]{Color-color diagram showing the basic A/B/C states and
	their observed transitions.}

\figcaption[]{RXTE/ASM light curve for the period analyzed in this paper.
	Bin size is 1 day. The bands in the upper part of the plot show
	the position of the quiet intervals examined in Section 3.4.}

\figcaption[]{Four examples of color-color diagrams (left) and light curves
	(right) for class $\chi$.}

\figcaption[]{Color-color diagram with the average values for observations in
	class $\chi$, subdivided by sub-class (see text). Average points
	for observations in class $\phi$ are also shown as diamonds. Circles
	are the same circles in Fig. 3c. The lines show the expected positions
	of theoretical models (dbb=disk-blackbody, pl= power law) for different
	values of N$_H$.}

\figcaption[]{Panels a,b: CD and light curve of an example of class $\phi$
	respectively. Panels c,d: CD and light curve of an example
	of class $\delta$ respectively. Points in panel c correspond only
	to the time intervals between the vertical lines, and are marked
	correspondingly. The arrows in panel d indicate the position of the
	state A transitions discussed in the text.}

\figcaption[]{Sample observation from class $\mu$. Panels a,b,c contain
	only part of the data, while panel d shows the whole dataset.
	(a) Color-intensity diagram; (b) light curve; (c)
	color-color diagram; (d) complete light curve. The symbols are the
	same as in Fig. 3}

\figcaption[]{Panel a: color-color diagram. Circles and diamonds are the same
	as in Fig. 10. Crosses are dips to state A (see text). The dot-dash
	line corresponds to a disk-blackbody model with N$_H$ fixed to
	6$\times 10^{22}$cm$^{-2}$. Panel b: same points, but in a
	color-intensity diagram.}

\figcaption[]{Sample observation from class $\gamma$. Panels a,b,c contain
	only part of the data, while panel d shows the whole dataset.
	(a) Color-intensity diagram; (b) light curve; (c)
	color-color diagram; (d) complete light curve. The symbols are the
	same as in Fig. 3}

\figcaption[]{Sample observation from class $\rho$. Panels a,b,c contain
	only part of the data, while panel d shows the whole dataset.
	(a) Color-intensity diagram; (b) light curve; (c)
	color-color diagram; (d) complete light curve. The symbols are the
	same as in Fig. 3}

\figcaption[]{Two examples of light curves at high (1/8 sec) time resolution
	light curves from observation K-03-00, shifted so that the small
	peaks at the end of the event match in time.}

\figcaption[]{Sample observation from class $\nu$. Panels a,b,c contain
	only part of the data, while panel d shows the whole dataset.
	(a) Color-intensity diagram; (b) light curve; (c)
	color-color diagram; (d) complete light curve. The symbols are the
	same as in Fig. 3}

\figcaption[]{Sample observation from class $\alpha$. Panels a,b,c contain
	only part of the data, while panel d shows the whole dataset.
	(a) Color-intensity diagram; (b) light curve; (c)
	color-color diagram; (d) complete light curve. The symbols are the
	same as in Fig. 3. Panel e shows an enlargement of the biggest
	peak, and panel f shows the corresponding CD.}

\figcaption[]{Sample observation from class $\alpha$. Panels a,b,c contain
	only part of the data, while panel d shows the whole dataset.
	(a) Color-intensity diagram; (b) light curve; (c)
	color-color diagram; (d) complete light curve. The symbols are the
	same as in Fig. 3. Panel e shows an enlargement of an oscillating
	section of the light curve.}

\clearpage

\begin{deluxetable}{lcl}
\footnotesize
\tablecaption{The observation IDs of the pointing analyzed in this work.
	The letters I,J,K stand for 10408-01, 20187-02, and 20402-01
	respectively. The second column indicates how many observation
	intervals are included in this class. The \# indicates the
	individual observation interval.
	When no \# is specified, all intervals are included.}
\tablewidth{0pt}
\tablehead{
  \colhead{Class} &
  \colhead{N.} &
  \colhead{Observation IDs} \nl
}
\startdata

$\alpha$ & 23 &J-01-00,J-01-01,J-02-00(\#2,\#3,\#5),K-22-00,K-23-00,
	   K-24-01,K-26-00(\#3,\#4),K-27-00,K-28-00, \nl
	 && K-30-01(\#2), K-30-02 \nl

$\beta$	 & 25 &I-10-00,I-21-00,I-21-01(\#1),K-43-00(\#2),K-43-02,
	   K-44-00,K-45-00(\#1,\#2,\#3,\#5),K-45-03, \nl
	 && K-46-00,K-52-01,K-52-02,K-53-00,K-59-00(\#2); \nl

$\gamma$ & 22 &I-07-00(\#2,\#3),K-37-00,K-37-02,K-38-00,
	   K-39-00,K-39-02,K-40-00,K-55-00,K-56-00, K-57-00 \nl

$\delta$ & 33 &I-13-00(\#1),I-14-00,I-14-01,I-14-02,I-14-03,
	   I-14-04,I-14-05,I-14-07,I-14-08,I-14-09,I-17-00, \nl
	 && I-17-03,I-18-00(\#1,\#2),I-18-01,I-18-04,K-41-00,
	   K-41-01,K-41-02(\#1),K-41-03,K-42-00,K-53-02(\#2), \nl
	 &&K-54-00 \nl

$\theta$ & 17 &I-15-00,I-15-01,I-15-02,I-15-03,I-15-04,I-15-05,
	   I-16-00,I-16-01,I-16-02,I-16-03,I-16-04, \nl
	 && I-21-01(\#2),K-45-02 \nl

$\kappa$ & 4  &K-33-00,K-35-00 \nl

$\lambda$& 10 &I-37-00(\#1,\#2),
	   I-38-00(\#3,\#4,\#5),K-36-00,K-36-01,K-37-01 \nl

$\mu$	 & 18 &I-08-00,I-34-00,I-35-00,I-36-00,
	   K-43-00(\#1),K-43-01,K-45-01,
	   K-53-01,K-53-02(\#1),K-59-00(\#3) \nl

$\nu$	 & 18 &I-37-00(\#3),I-40-00,I-41-00,
	   I-44-00,K-01-00(\#3),K-02-02(\#4) \nl

$\rho$	 & 19 &K-03-00,K-27-02,K-30-00,K-31-00,
	   K-31-01,K-31-02,K-32-00,K-32-01,K-34-00,
	   K-34-01 \nl

$\phi$	 & 18 &I-09-00,I-11-00,I-12-00,I-13-00(\#2,\#3),
	   I-17-01,I-17-02,I-18-00(\#3),I-19-00,I-19-01,I-19-02, \nl
	 && I-20-00,I-20-01 \nl

$\chi_1$ & 30 &I-22-00,I-22-01,I-22-02,I-23-00,I-24-00,I-25-00,
	   I-27-00,I-28-00,I-29-00,I-30-00,I-31-00\nl

$\chi_2$ & 47 &K-04-00,K-05-00,K-07-00,K-08-00,K-08-01,K-09-00,K-10-00,
	   K-11-00,K-12-00,K-13-00,K-14-00,K-15-00, \nl
	 && K-16-00,K-17-00,K-18-00,K-19-00,K-20-00,K-21-01, \nl

$\chi_3$ & 15 &K-49-00,K-49-01,K-50-00,k-50-01,K-51-00,K-52-00 \nl

$\chi_4$ & 49 &I-31-00,I-32-00,I-33-00,I-38-00(\#2),I-42-00,I-43-00,I-45-00,
	   J-02-00(\#1,\#4,\#6),K-01-00(\#1,\#2),\nl
	&& K-02-02(\#1,\#2,\#3),K-24-00,K-25-00,K-26-00(\#1,\#2,\#5),
	     K-26-01,K-26-02,K-27-01,K-27-03,K-29-00,\nl
	  && K-30-01(\#1),K48-00 \nl
\enddata
\end{deluxetable}

\clearpage
\centerline{\psfig{figure=f1.ps,width=15cm} {\hfil}}
\centerline{\it Figure 1}
\vskip 0.5 true cm
\clearpage
\centerline{\psfig{figure=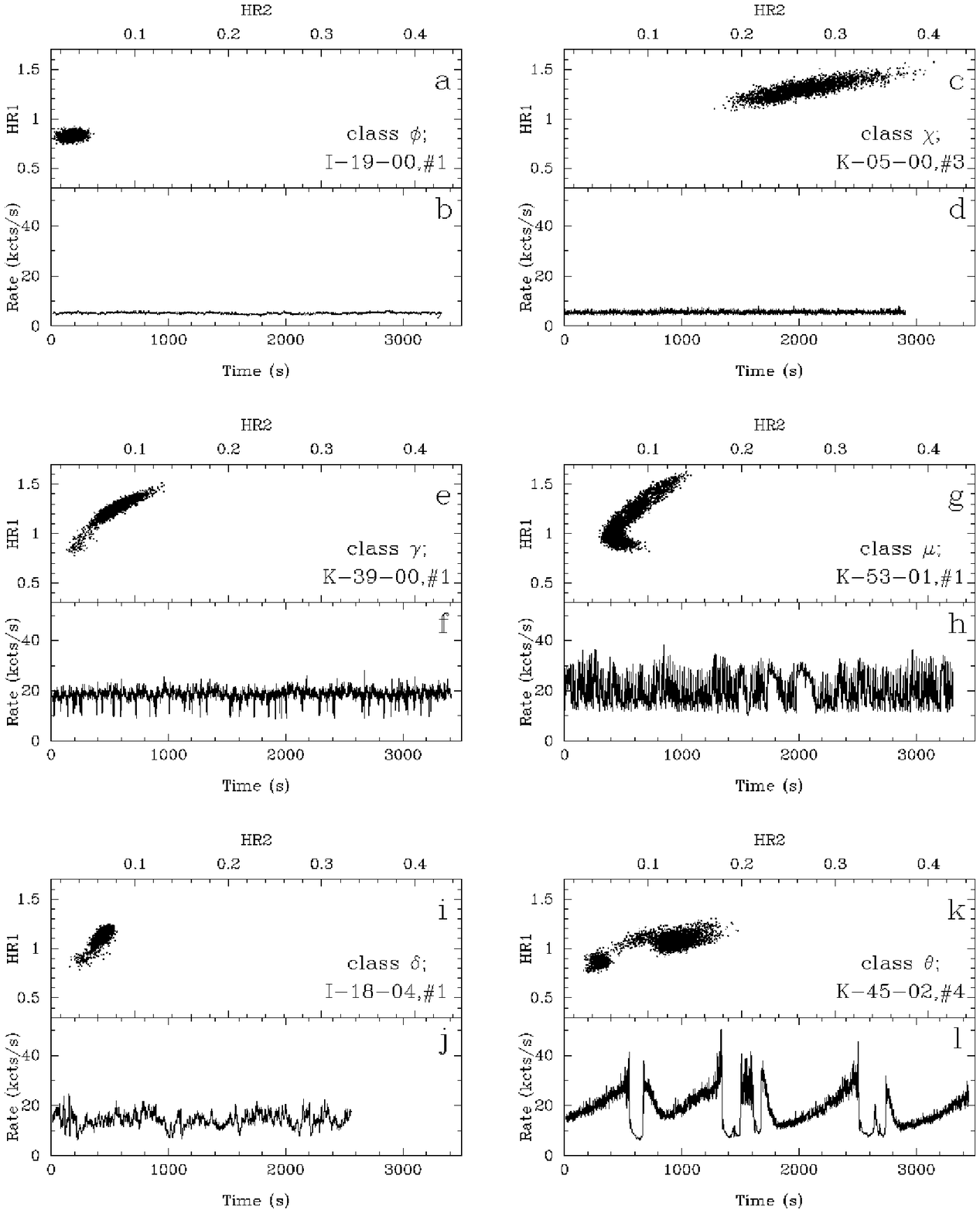,width=15cm} {\hfil}}
\centerline{\it Figure 2}
\vskip 0.5 true cm
\clearpage
\centerline{\psfig{figure=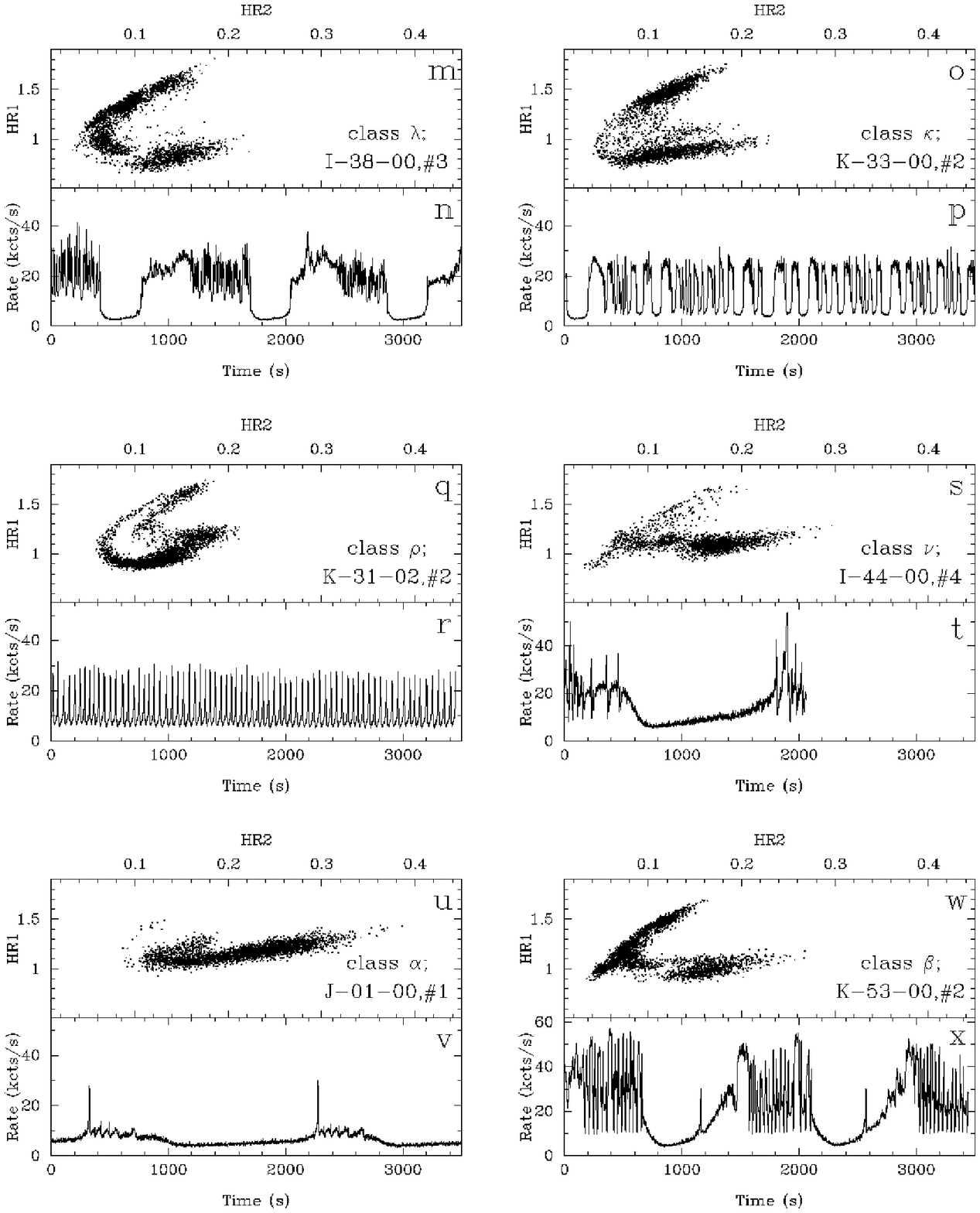,width=15cm} {\hfil}}
\centerline{\it Figure 2 (cont'd)}
\vskip 0.5 true cm
\clearpage
\centerline{\psfig{figure=f3.ps,width=15cm} {\hfil}}
\centerline{\it Figure 3}
\vskip 0.5 true cm
\clearpage
\centerline{\psfig{figure=f4.ps,width=15cm} {\hfil}}
\centerline{\it Figure 4}
\vskip 0.5 true cm
\clearpage
\centerline{\psfig{figure=f5.ps,width=15cm} {\hfil}}
\centerline{\it Figure 5}
\vskip 0.5 true cm
\clearpage
\centerline{\psfig{figure=f6.ps,width=15cm} {\hfil}}
\centerline{\it Figure 6}
\vskip 0.5 true cm
\clearpage
\centerline{\psfig{figure=f7.ps,width=15cm} {\hfil}}
\centerline{\it Figure 7}
\vskip 0.5 true cm
\clearpage
\centerline{\psfig{figure=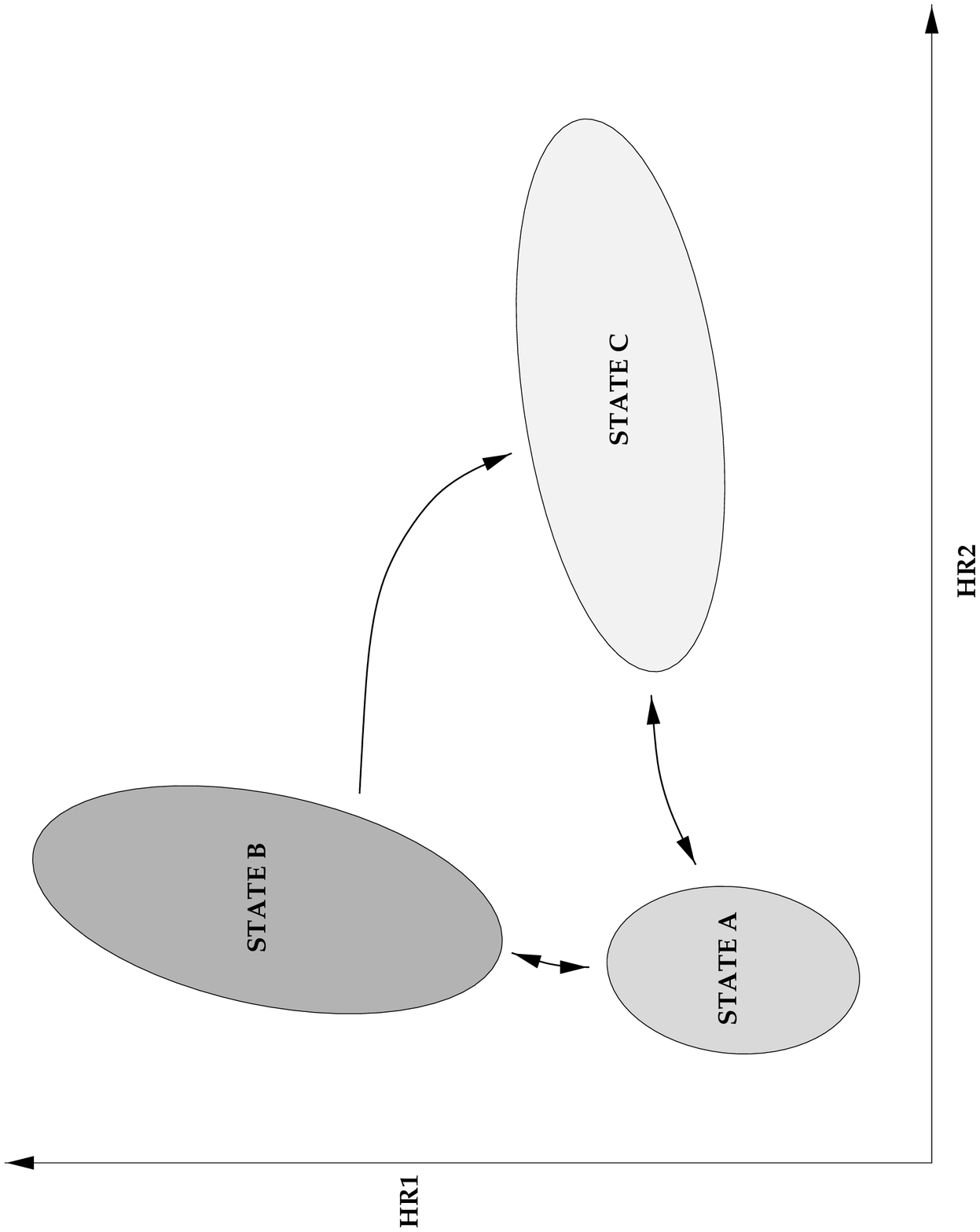,width=15cm} {\hfil}}
\centerline{\it Figure 8}
\vskip 0.5 true cm
\clearpage
\centerline{\psfig{figure=f9.ps,width=15cm} {\hfil}}
\centerline{\it Figure 9}
\vskip 0.5 true cm
\clearpage
\centerline{\psfig{figure=f10.ps,width=13cm} {\hfil}}
\centerline{\it Figure 10}
\vskip 0.5 true cm
\clearpage
\centerline{\psfig{figure=f11.ps,width=15cm} {\hfil}}
\centerline{\it Figure 11}
\vskip 0.5 true cm
\clearpage
\centerline{\psfig{figure=f12.ps,width=15cm} {\hfil}}
\centerline{\it Figure 12}
\vskip 0.5 true cm
\clearpage
\centerline{\psfig{figure=f13.ps,width=15cm} {\hfil}}
\centerline{\it Figure 13}
\vskip 0.5 true cm
\clearpage
\centerline{\psfig{figure=f14.ps,width=15cm} {\hfil}}
\centerline{\it Figure 14}
\vskip 0.5 true cm
\clearpage
\centerline{\psfig{figure=f15.ps,width=15cm} {\hfil}}
\centerline{\it Figure 15}
\vskip 0.5 true cm
\clearpage
\centerline{\psfig{figure=f16.ps,width=15cm} {\hfil}}
\centerline{\it Figure 16}
\vskip 0.5 true cm
\clearpage
\centerline{\psfig{figure=f17.ps,width=15cm} {\hfil}}
\centerline{\it Figure 17}
\vskip 0.5 true cm
\clearpage
\centerline{\psfig{figure=f18.ps,width=15cm} {\hfil}}
\centerline{\it Figure 18}
\vskip 0.5 true cm
\clearpage
\centerline{\psfig{figure=f19.ps,width=15cm} {\hfil}}
\centerline{\it Figure 19}
\vskip 0.5 true cm
\clearpage
\centerline{\psfig{figure=f20.ps,width=15cm} {\hfil}}
\centerline{\it Figure 20}
\vskip 0.5 true cm

\end{document}